# Structure And Properties of Nanoparticles Formed under Conditions of Wire Electrical Explosion

Y.S. Kwon[1], A.P. Ilyin[2], D.V. Tikhonov[2], V.V. An[2], A. Yu. Godimchuk[2], L.O. Tolbanova[2]

1- Research Center for Machine Parts and Materials Processing, School of Materials and Metallurgical Engineering, University of Ulsan, P.O. Box 18, San-29, Mugeo-Dong, Nam-Ku, Ulsan 680-749, South Korea

2- High Voltage Research Institute, Tomsk Polytechnic University, 2a Lenin Ave., Tomsk 634050, Russia

Wire electrical explosion (WEE) occurs when a power current pulse passes through a wire ($j \geq 10^{14}$ A/m$^2$). During the fast explosion metal (wire) proceeds to a non-equilibrium state which is characterized by two types of non-equilibrium. Firstly, when WEE primary products form, the electric energy leads to the electron subsystem excitation – staying "colder" (~$10^4$ K). Secondly, the WEE primary products are: vapor (clusters), plasma and overheated liquid drops, which cannot exist together under equilibrium conditions. At the same time, the energy input during WEE is equal to 1-2 sublimation energies, but 30 % of wire material can pass to the ion state because of non-equilibrium. At that, the wire magnetic field has a compression effect on the WEE primary products. Electric and magnetic fields have an impact on the formation of WEE secondary products, providing discontinuity of expanding products front and appearance of turbulences. The WEE secondary products are a system liquid (drops) − vapor (clusters). Their formation occurs during cooling (T<5000 K) due to condensation of vapors on charged centers of condensation [1].

The objective of this work is to establish micro- and substructure features of particles, a link between these features and properties of the nanopowders produced by WEE.

**Microstructure of nanoparticles.** The average cooling rate of nanoparticles during WEE is more than $10^7$ K/s, but the temperature decreases exponentially with time. According to our estimations, the cooling rate of particles does not exceed $1 \cdot 10^3$ K for the 5000-800 K range. The particles in this range are in a liquid state for a relatively long time and subjected to Laplace's compression (~$10^7$-$10^8$ Pa).

It was experimentally shown that aluminum nanopowders produced in gaseous argon can be divided in an electrofilter by the charge before contacting air. It follows that the surface of particles is protected by a non-conducting layer ("argon coat") and such particles hold the charge. The positive charge on the surface of particles increases during passivation. Namely, the Laplace's compression and the charge of the surface determine the spherical shape of particles in the liquid and the solid state. Passivation of copper and aluminum nanoparticles with a characteristic size of ~3-40 nm is accompanied by their oxidation and crystallization.

**Substructure of nanoparticles.** X-ray structure analysis has shown that there is no polycluster structure of nanoparticles. Inside of particles, there are no gas inclusions, linear and volume defects. Obviously, they were pressed out top the surface under Laplace's compression where they had recombined. The substructure specific for nanoparticles of metals, alloys and chemical compounds is static displacements in the crystal lattice from their position in the equilibrium state. Such a kind of deficiency is characteristic for all types of nanopowders produced by WEE. Apparently, it can be explained by a relatively fast transition from the liquid state to the solid one. Probably, the process of nanoparticles solidification occurs from the surface. First, a solid product has a compressing effect on the liquid particle, and then a tensile effect takes a place. Indeed for iron particles of 0,5-1 μm, hollow particles form during WEE. If the WEE products have several polymorph states of the crystal lattice, a lattice with a minimal X-ray density stabilizes.

In the literature, the interpretation of data on low-angled X-ray scattering is still under discussion. Usually, an observed reflection is linked to defect-free areas of X-ray coherent scattering ($D_{ACS}$). It was experimentally shown that $D_{ACS}$ sizes decrease with a decrease of particle diameter. Besides, the thickness of a protective oxide-hydroxide shell on particles decreases also with a decrease of particle diameter. Direct structure observations of a nanoparticles section using high resolution electron microscope (0,1 Å) have shown that there are no grain boundaries and severe changes in the substance density. At the same time, a correlation between the thickness of the oxide-hydroxide layer and X-ray analysis data is observed.

Characteristic properties of nanopowders produced by WEE are: heightened sintering and oxidation resistance at normal temperatures, processes passing at heating under threshold phenomena regimes. These properties can be explained by relaxation processes starting to occur in the structure of WEE powders in a narrow temperature range (fig.1).

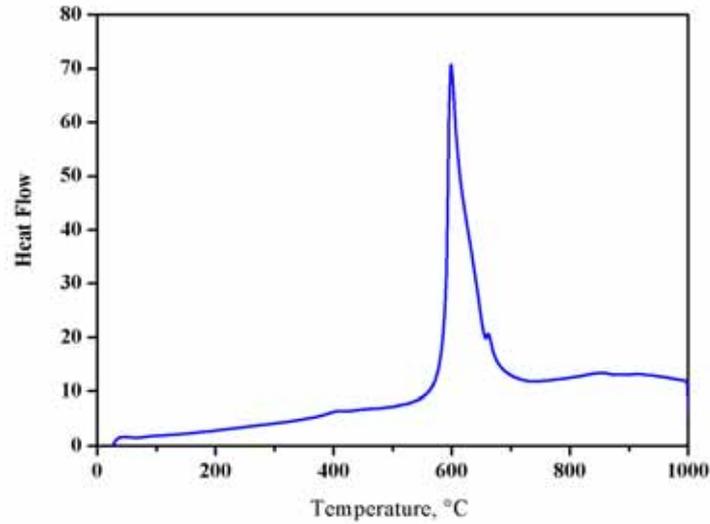

Fig.1. DTA trace of aluminum nanopowder.

Simultaneous occurrence of relaxation processes and oxidation leads to a heat release in the zone of interaction. In case of the bad thermal conductivity, these processes occur under conditions of a thermal explosion. When sintering nanopowders without access for air, the main source of heat release is a recombination of charges in the double electric layer. This layer is formed due to oxidation-reduction processes and is characterized by a high pseudocapacity (fig.2). A thermal breakdown of the double electric layer determines indeed an onset of self-heating and sintering.

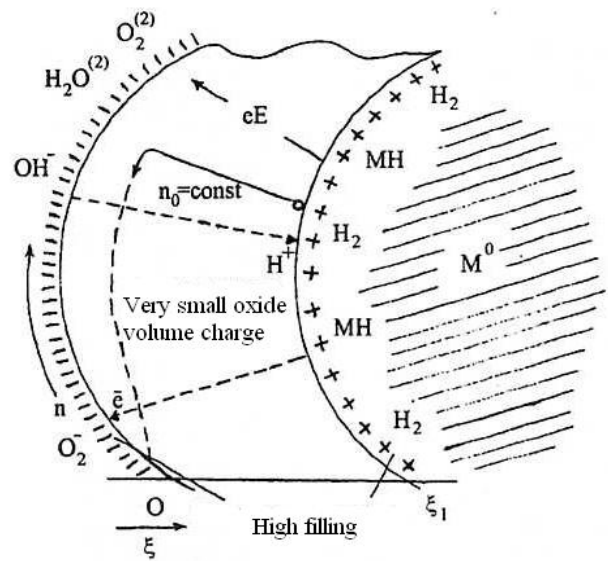

Fig.2. Scheme of the double electric layer in the nanosized particle.

In the presence of an oxidizer, e.g., during heating of nanopowders in air, the oxidation onset temperature does not depend on fabrication conditions and their dispersivity in studied ranges of particle sizes. Results of derivatogram processing for a series of aluminum nanopowders are given in Tab.1.

Tab.1. Results of derivatogram processing for aluminum nanopowders

| Nanopowder | $S_{sp}$, m$^2$/g | [Al$^0$], wt% | Oxidation onset T$^0$, °C |
|---|---|---|---|
| Al-1 | 10,8 | 88,0 | 540 |
| Al-2 | 9,9 | 87,9 | 530 |
| Al-3 | 9,8 | 88,1 | 550 |
| Al-4 | 9,3 | 88,5 | 540 |
| Al-5 | 8,8 | 90,9 | 550 |
| Al-6 | 6,7 | 90,0 | 540 |
| Al-7 | 7,7 | 91,0 | 550 |

Aluminum nanopowders have been produced under different conditions of wires electrical explosion in argon. The energy input into the exploded wires was 9.9 to 2.1 of sublimation energy of wires. Apart from samples 6 and 7, the fabrication of other nanopowders was accompanied by ignition of an electric arc in WEE products. Nevertheless, the obtained results testify that the fabrication conditions of aluminum nanopowders do not practically influence the oxidation onset temperature when heating them in air.

## Conclusion

The metal nanopowders obtained under wires electrical explosion are characterized by a spherical shape of particles. Such a shape of particles shows their stabilization through a stage of liquid state. A specific type of deficiency is characteristic for the particles – static displacements of atoms from the equilibrium state. On the whole, the state of WEE powder particles can be characterized as a metastable state. It leads to an increased stability of nanopowders at normal temperatures and an increased reactivity during heating, which is revealed in the form of threshold phenomena.